\documentstyle[aps,prd,epsf,floats]{revtex} 

\draft

\begin{document}
\twocolumn[\hsize\textwidth\columnwidth\hsize\csname
@twocolumnfalse\endcsname 

\begin{flushright}
UCLA/98/TEP/30
\end{flushright}

\preprint{UCLA/98/TEP/30}
\title{Pulsar kicks from neutrino oscillations}

\author{Alexander Kusenko} 
\address{Department of Physics and Astronomy, UCLA, Los Angeles, CA
90095-1547 } 

\author{Gino Segr\`e}
\address{Department of Physics, University of Pennsylvania, Philadelphia, 
PA 19104}

\date{November, 1998}

\maketitle
             
\begin{abstract}
Neutrino oscillations can explain the observed motion of 
pulsars.  We show that two different models of neutrino emission
from a cooling neutron star are in good quantitative agreement and predict
the same order of magnitude for the pulsar kick velocity, consistent with the
data.   
\end{abstract}


\vskip2.0pc]


\renewcommand{\thefootnote}{\arabic{footnote}}
\setcounter{footnote}{0}

\section{Introduction}
\label{sec-1}

We recently suggested~\cite{ks96,ks97} that the observed proper motions of
pulsars~\cite{astro} may be the result of neutrino oscillations in the hot
neutron star born in a supernova explosion.  Neutrinos are the only
significant cooling agents during the first 10 seconds after the onset of
the supernova, and they carry away most of the energy liberated in the
gravitational collapse, $\sim 10^{53}$erg.  A 1\% asymmetry in the
distribution of neutrinos can account for the measured pulsar velocities 
$\sim 500$ km/s. 

Neutrino oscillations in medium are affected by the polarization effects
due to magnetic field~\cite{polar}.  Consequently, the strong magnetic
field of a neutron star can affect the depth at which the neutrino
conversions take place.  The points of the Mikheev-Smirnov-Wolfenstein
resonance lie on a surface that is not concentric with the star.  For a
certain range of neutrino masses, this
creates an asymmetry in the distribution of momenta of the outgoing
neutrinos.  The $\nu_{\mu,\tau} \leftrightarrow \nu_{e}$
oscillations~\cite{ks96}, as well as the sterile-to-active neutrino
conversions~\cite{ks97}, can give the pulsar a kick consistent with
the observation provided that the magnetic field inside the star is of order
$10^{14} - 10^{15}$G.  Other types of neutrino conversions could also
produce a similar effect~\cite{others}. 

Our explanation~\cite{ks96,ks97} uses the fact that different neutrino
species have different opacities in nuclear matter.  The charged-current
interactions are responsible for the difference in the mean free paths of
$\nu_e$ and $\nu_{\mu,\tau}$.  They are also the cause of
the matter-enhanced $\nu_e \leftrightarrow \nu_{\mu,\tau}$ conversions. 
In the case of sterile neutrinos, both charged and neutral currents
contribute to the difference in the opacities.   
The analyses of Refs.~\cite{ks96,ks97,others} was based on a 
simplified model which, as we will see, gave the right order-of-magnitude
estimate for the pulsar velocity.  

Recently, a different model for the neutrino emission was used to calculate
the kick~\cite{jr} due to the active neutrino 
oscillations\footnote{The active-to-sterile neutrino conversions of
Ref.~\cite{ks97} are assumed to take place at much higher densities than
those discussed in Ref.~\cite{jr}.}.  The result obtained in Ref.~\cite{jr}
was incorrect because the neutrino absorption $\nu_e n \rightarrow e^-p^+$ 
was neglected and also because the different neutrino opacities were
assumed to be equal to each other.  We emphasize that in the absence of
charged-current interactions the kick from the active neutrino
oscillations~\cite{ks96} should vanish.  

We will show that, after the charged-current interactions are included, the
two models are, in fact, in good agreement, as they should be. 

\section{Hard neutrinospheres}

Our calculations \cite{ks96,ks97} employed a model with a sharp
neutrinosphere, such that the neutrinos of a given type were assumed
trapped inside and free-streaming outside.  Several subsequent analyses 
\cite{others} used the same model to calculate the kick to the pulsar from
the neutrinos whose opacities changed on their passage through matter due
to oscillations. 
 
In this model one assumes that the luminosity of the emitted neutrinos 
obeys the Stefan-Boltzmann law and is proportional to $T^4$. 
 
This model is admittedly simplistic.  However, it appears to work well
insofar as predicting the order of magnitude of the kick.

\section{Soft neutrinospheres}

Let us now consider the Eddington model for the atmosphere which was used
by Schinder and Shapiro~\cite{ss} to describe the emission of a single
neutrino species.  We will generalize it to include several types of
neutrinos. 

In the diffusion approximation, the distribution functions $f$ are taken in
the form~\cite{ss}:   
\begin{equation}
f_{\nu_i} \approx f_{\bar{\nu}_i}
\approx f^{eq} + \frac{\xi}{\Lambda_i} \frac{\partial
f^{eq}}{\partial m}, 
\label{f_diff}
\end{equation}
where $f^{eq}$ is the distribution function in equilibrium, $\Lambda_i$
denote the respective opacities, $m$ is the column mass density, $m=\int
\rho \: dx$, $\xi=cos \alpha$, and $\alpha $ is the normal angle of 
the neutrino velocity to the surface. 
At the surface, one imposes the same boundary condition for all the
distribution functions, namely 
\begin{equation}
\begin{array}{ll}
f_{\nu_i}(m,\xi)= 0, &  {\rm for} \ \xi < 0, \\
f_{\nu_i}(m,\xi)= 2 f^{eq}, &  {\rm for} \ \xi > 0.
\end{array} 
\label{bc}
\end{equation}
However, the differences in $\Lambda_i$ produce unequal distributions for
different neutrino types. 

Generalizing the discussion of Refs.~\cite{jr,ss} to include six flavors,
three neutrinos and three antineutrinos, one can write the energy flux as 

\begin{equation}
F=2\pi \int_0^\infty E^3 dE \int_{-1}^1 \xi d \xi \ 
\sum_{i=1}^{3} (f_{\nu_i}+f_{\bar{\nu}_i}),
\end{equation}
We will assume that $\Lambda_i=\Lambda_i^{(0)} (E^2/E^2_0)$.  

We use the expressions for $f_{\nu_i}$ from equation (\ref{f_diff}). 
Changing the order of differentiation with
respect to $m$ and integration over $E$ and $\xi$, and 
using the fact that $f^{eq}$ is isotropic, we arrive at
the result similar to that of Ref.~\cite{ss}: 

\begin{equation}
F=\frac{2\pi^3}{9} E_0^2  \left [ \sum_{i=1}^{3} \frac{2}{\Lambda^{(0)}_i} 
\right ] 
\frac{\partial T^2}{\partial m}.
\end{equation}

The basic assumption of the model is that flux $F$ is conserved.  In other
words, the neutrino absorptions   $\nu_e n \rightarrow e^- p^+$ are
neglected.  Since the sum in brackets, as well as the flux $F$ are
treated~\cite{ss} as constants with respect to $m$, one can solve
for $T^2$:  

\begin{equation}
T^2(m) = \frac{9}{2\pi^3} E_0^{-2} \left [ \sum_{i=1}^{3}
\frac{2}{\Lambda^{(0)}_i}  \right ]^{-1} F \, m+ \left(\frac{30}{7\pi^5} F
\right )^{1/2} 
\label{t2}
\end{equation}

Swapping the two flavors in equation (\ref{t2}) leaves the
temperature unchanged in the Eddington approximation.  Hence, neutrino
oscillations do not alter the temperature profile in this
approximation\footnote{ It was also assumed in our earlier 
calculations~\cite{ks96,ks97} that neutrino oscillations do not have a 
significant effect on the temperature profile.  We then took the spectral
temperature of free-streaming neutrinos to be equal to matter temperature
at their respective neutrinospheres.}.  

We will now include the absorptions of neutrinos. 

Some of the electron neutrinos are absorbed on their passage through the
atmosphere thanks to the charged-current process

\begin{equation}
\nu_e n \rightarrow e^- p^+.
\end{equation}
The cross section for this reaction is $\sigma = 1.8 \: G_{_F}^2
E_\nu^2 $, where $E_\nu$ is the neutrino energy.  The total momentum
transfered to the neutron star by the passing neutrinos depends on the
energy. 

Both numerical and analytical calculations show that the muon
and tau neutrinos leaving the core have much higher mean energies than the
electron neutrinos~\cite{suzuki,energies}.  Below the point of MSW resonance
the electron neutrinos have the mean energies $\approx 10$~MeV, while the muon
and tau neutrinos have energies $\approx 25$~MeV.  

The origin of the kick in this description is that the neutrinos spend more
time as energetic electron neutrinos on one side of the star than on the other
side, hence creating the asymmetry.  Although the temperature profile
remains unchanged in Eddington approximation, the unequal numbers of
neutrino absorptions push the star, so that the total momentum is
conserved. 

Below the resonance
$E_{\nu_e}<E_{\nu_{\tau,\mu}}$. Above the resonance, this relation is
inverted.  The energy deposition into the nuclear matter depends on the
distance the electron neutrino has traveled with a higher energy.  This
distance is affected by the direction of the magnetic field relative to the
neutrino momentum.

We assume that the resonant conversion $\nu_e \leftrightarrow \nu_\tau$
takes place at the point $r=r_0+\delta(\phi); \delta (\phi)=\delta_0 \cos
\phi$.  The position of the resonance depends on the
magnetic field $B$ inside the star~\cite{ks96}: 

\begin{equation}
\delta_0=\frac{e \mu_e B}{2 \pi^2} \left / \frac{dN_e}{dr} \right., 
\end{equation}
where $N_e=Y_e N_n$ is the electron density and $\mu_e$ is the electron
chemical potential. 

Below the resonance the  
$\tau$ neutrinos are more energetic than the electron neutrinos.  The
oscillations exchange the neutrino flavors, so that above  the resonance
the electron neutrinos are more energetic than the $\tau$ neutrinos.  The
number of neutrino absorptions in the layer of thickness $2 \delta (\phi)$
around $r_0$ depends on the angle $\phi$ between the neutrino momentum and
the direction of the magnetic field.  Each occurrence of the neutrino
absorption transfers the momentum $E_{\nu_e}$ to the star.  The difference
in the numbers of collisions per electron neutrino between the directions
$\phi$ and $-\phi$ is  

\begin{eqnarray}
\Delta k_e/E_{\nu_e} & = & 2 \: \delta(\phi) \: 
N_n \: [\sigma(E_1)-\sigma(E_2)] \\
& = & 1.8 \: G_{_F}^2 [E_1^2-E_2^2] \: \frac{\mu_e}{Y_e} \: \frac{eB}{\pi^2} 
\: h_{N_e} \:  \cos \phi,
\end{eqnarray}
where $h_{N_e}=[d(\ln N_e)/dr]^{-1}$. 

We use $Y_e\approx 0.1$, $E_1\approx 25$~MeV, $E_2\approx10$~MeV,
$\mu_e\approx 50$~MeV, and $h_{N_e}\approx 6$~km. 
After integrating over angles and taking into account that only one
neutrino species undergoes the conversion, we obtain the final result for the
asymmetry in the momentum deposited by the neutrinos: 

\begin{equation}
\frac{\Delta k}{k} = 0.01 \frac{B}{2\times 10^{14} {\rm G}},
\label{final} 
\end{equation}
which agrees with the earlier estimates\footnote{We note in passing that we
estimated the kick in Refs.~\cite{ks96,ks97} assuming $\mu_e \approx
const$. A different approximation, $Y_e \approx const$, gives a somewhat
higher prediction for the magnitude of the magnetic field~\cite{comment}. 
}~\cite{ks96,comment}. 

Neutrinos also lose energy by scattering off the electrons.  Since the
electrons are degenerate, the final-state electron must have energy greater
than $\mu_e$.  Therefore, electron neutrinos lose from $0.2$ to $0.5$ of
their energy per collision in the neutrino-electron scattering.  However,
since $N_e \ll N_n $, this process can be neglected. 

One may worry whether the asymmetric absorption can produce some
back-reaction and change the temperature distribution inside the star 
altering our result (\ref{final}).  If such effect exists, it is 
beyond the scope of Eddington approximation, as is clear from equation
(\ref{t2}).  The only effect of the asymmetric absorption is to
make the star itself move, in accordance with the momentum conservation.
This is the origin of the kick (\ref{final}). 

Of course, in reality the back-reaction is not exactly zero.  The most
serious drawback of Eddington model, pointed out in Ref.~\cite{ss}, is that
diffusion approximation breaks down in the region of interest, where the
neutrinos are weakly interacting.  Another problem has to do with
inclusion of neutrino absorptions and neutrino oscillations~\cite{ss}. 
However, to the extent we believe this
approximation, the pulsar kick is given by equation~(\ref{final}). 

\section{Comparison of the two models}

We have juxtaposed two models for neutrino transfer.  One of them assumes
that neutrinos are trapped inside a sharply defined neutrinosphere, but that
they are free-streaming outside.   In this model, the
absorption of neutrinos outside the neutrinosphere is neglected and the
asymmetry arises from the change in the position of the neutrinosphere due
to neutrino oscillations.

The other model considers the propagation of the neutrinos through the
atmosphere interpolating between the region where the neutrinos are trapped
and that in which they are free-streaming.  Here the neutron star receives
a kick from the unequal numbers of reactions $\nu_e n \rightarrow p^+ e^-$
on either side. The higher-energy neutrinos spend more time as electron
neutrinos on one side of the star than on the other side.  Therefore, the
passage of neutrinos creates an uneven drag to the star. 

Incidentally, the Stefan-Boltzmann relation between the luminosity and
temperature is, of course, present in both models.  In the diffusion
approximation, equation (\ref{t2}) implies that $F \propto T^4$ at $m=0$. 

In both models the kick is a manifestation of the unequal neutrino
opacities which are caused by the charged-current interactions. 

The two models are in good quantitative agreement.

\section{Conclusion}

The neutrino oscillations can be the explanation of the pulsar motions.
Although many alternatives have been examined, all of them fail
to explain the large magnitudes of the pulsar velocities.  Parity violation
effects recently claimed to explain the kicks~\cite{cumulative} turned out to
be irrelevant because they wash out in statistical equilibrium and produce no
appreciable asymmetry in the neutrino luminosity~\cite{eq}.  Other proposed
mechanisms required unusually high neutrino magnetic
moments~\cite{voloshin} or some exotic interactions.  The oscillations
appear to be the only viable explanation at present. 

If the pulsar kick velocities are due to $\nu_e \leftrightarrow
\nu_{\mu,\tau} $ conversions, one of the neutrinos must have mass $\sim 
100$~eV (assuming small mixing) and must decay on the cosmological time
scales not to overclose the Universe~\cite{ks96}.  This has profound 
implications for particle physics hinting at 
the existence of Majorons~\cite{peccei} or other physics beyond the
Standard Model that can facilitate the neutrino decay. 

If the active-to-sterile neutrino oscillations~\cite{ks97} are responsible
for pulsar velocities, the prediction for the sterile neutrino to have a mass
of several keV is not in contradiction with any of the present bounds.  In
fact, the $\sim$keV mass sterile neutrino has been proposed as a
dark-matter candidate~\cite{fs}. 

To summarize, two different models of neutrino emission predict
the same order of magnitude for the pulsar birth velocity caused by
neutrino oscillations in the cooling neutron star.  


\end{document}